\documentclass{jkas}

\def\year{2015} 
\def\month{April} 
\def\volume{48} 
\def\issue{2} 
\def\beginpage{1} 
\def\endpage{13} 
\def\received{February 9, 2015} 
\def\accepted{March 16, 2015} 
\date{Received \received; accepted \accepted}

\usepackage{flushend} 

\input colordvi

\title{
A New Hardware Correlator in Korea:\\ Performance Evaluation using
KVN observations
}

\author[1,2]{Sang-Sung Lee}
\author[1]{Chung Sik Oh}
\author[1]{Duk-Gyoo Roh}
\author[1]{Se-Jin Oh}
\author[1,2]{Jongsoo Kim}
\author[1]{Jae-Hwan Yeom}
\author[1]{Hyo Ryoung Kim}
\author[1]{Dong-Gyu Jung}
\author[1]{Do-Young Byun}
\author[1,2]{Taehyun Jung}
\author[3]{Noriyuki Kawaguchi}
\author[4]{Katsunori M. Shibata}
\author[1]{Kiyoaki Wajima}

\affil[1]{Korea Astronomy and Space Science Institute, 776 Daedeokdae-ro,
Yuseong-gu, Daejeon 305-348, Korea; \email{sslee@kasi.re.kr}}
\affil[2]{Korea University of Science and Technology, 176 Gajeong-dong, Yuseong-gu, Daejeon 305-350, Korea}
\affil[3]{Shanghai Observatory, Chinese Academy of Sciences,
80 Nandan Road, 200030 Shanghai, P.R. China}
\affil[4]{Mizusawa VLBI Observatory, National Astronomical Observatory of Japan, 2-21-1 Osawa, Mitaka, Tokyo 181-8588, Japan}

\def\corrauthor{
S.-S. Lee
}

\def\runningauthor{
Lee et al.
}

\def\runningtitle{
A New Hardware Correlator in Korea: Performance Evaluation with KVN
}

\def\keywords{
Techniques: interferometric --- Instrumentation: interferometers
--- Radio continuum: galaxies --- Masers
}

\def\abstracttext{
We report results of the performance evaluation of a new hardware correlator
in Korea, the Daejeon correlator, developed by the Korea Astronomy and Space Science
Institute (KASI) and the National Astronomical Observatory of Japan (NAOJ).
We conducted Very Long Baseline Interferometry (VLBI) observations at 22~GHz
with the Korean VLBI Network (KVN) in Korea and the VLBI Exploration of Radio Astrometry (VERA) in Japan,
and correlated the aquired data with the Daejeon correlator.
For evaluating the performance of the new hardware correlator,
we compared the correlation outputs from
the Daejeon correlator for KVN observations
with those from a software correlator, the Distributed FX (DiFX).
We investigated the correlated flux densities and
brightness distributions of extragalactic compact radio sources.
The comparison of the two correlator outputs show that they are consistent with
each other within $<8\%$,
which is comparable with the amplitude calibration uncertainties of
KVN observations at 22~GHz.
We also found that the 8\% difference in flux density is caused mainly by
(a) the difference in the way of fringe phase tracking
between the DiFX software correlator and the Daejeon hardware correlator,
and (b) an unusual pattern (a double-layer pattern) of the amplitude 
correlation output from the Daejeon correlator. 
The visibility amplitude loss by the double-layer pattern is as small as 3\%. 
We conclude that the new hardware correlator produces 
reasonable correlation outputs
for continuum observations,
which are consistent with the outputs from the DiFX software correlator.
}

\begin{document}
\jkashead 


\section{Introduction\label{sec:intro}}

Very Long Baseline Interferometry (VLBI) is
an astronomical observing technique developed in the 1960s for measuring 
the accurate position of compact radio sources
and obtaining their sky brightness distribution at high angular resolution
by detecting fringes of noise signals arriving at two, or more, radio
telescopes from the celestial compact radio sources~\citep{RH60,cla+67,mor+67}.
The noise signals arriving at each radio telescope are filtered,
down-converted, and digitally sampled with being accurately time-tagged
by a frequency standard located at each observatory.
The sampled signals are recorded using magnetic tape systems such as
the Mark I system~\citep{bar+67},
the Mark II system~\citep{cla73},
the Mark III system~\citep{rog+83},
the Mark IV system~\citep{whi93},
and 
the S2 system~\citep{wie+96},
or to hard disk systems such as
the Mark 5 system~\citep{whi02}
and 
the K5 system~\citep{kon+03}.
The recording media are shipped to a correlator center and
played back. At that moment,  
one of the main roles of the correlator 
is to detect a fringe between the two signals
from radio telescopes after correcting for
geometric and instrumental delays of the signals.
Pre-estimates of the delays, {\it apriori} data
are applied to the correlation of the signals
from the two radio telescopes (or from a baseline).
The correlation output in this stage is known
as the visibility.
The correlation output streams are formatted in
Flexible Image Transport System (FITS)
interferometry Data Interchange Convention~\citep{gre09}.
Through a post-correlation process, the visibilities
are used for measuring the accurate positions 
of celestial radio sources and recovering their
sky brightness distributions.
The post-correlation process is performed
generally with astronomical processing programs:
for example, the Astronomical Image Processing System (AIPS).

A number of VLBI correlators have been developed to detect the fringes
from VLBI observations. The VLBI correlators include hardware and software
correlators~\citep{bar+67,mor+67,nap+94,wil+96,cas99,car+99,hor+00,del+07,del+11}.
Technical progress in parallel computing and high-speed networks
enables the construction of software correlators at low price and
with short development period.
Especially software correlator such as the DiFX~\citep{del+07,del+11}
became popular because of its easy installation and extensive support
from its user groups.
The price per baseline and price per unit data bandwidth 
are almost the same in software correlators,
while the prices decrease significantly with the number of baseline
and data rate in hardware correlators.
Therefore, hardware correlators become attractive when the target VLBI array
consists of a large number of antennas with a wide bandwith system.
In addition, the electric power consumption of hardware correlators is much less
than that of software correlators, significantly reducing the total operation budget.
These are the main reasons why connected arrays,
which usually have more elements and wider bandwidth than VLBI arrays,
still prefer hardware correlators.

\begin{figure}[t!]
        \centering 
        \includegraphics[width=83mm]{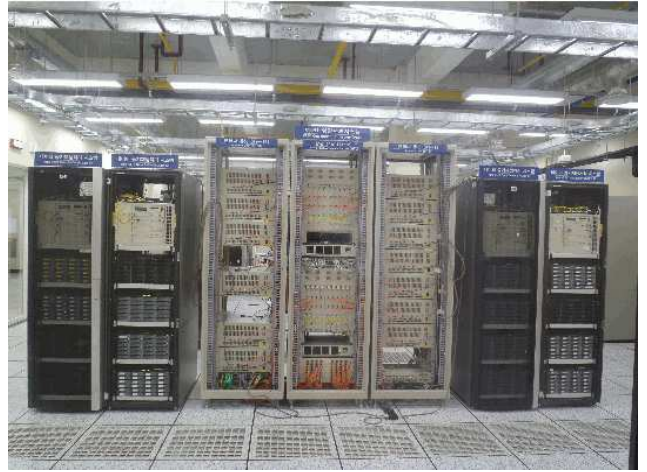}
        \caption{ 
        Daejeon correlator (KASI, Daejeon). Adopted from \cite{lee+14}
        }
        \label{fig1} 
\end{figure}

\begin{figure*}[!t]
        \centering 
        \includegraphics[width=170mm]{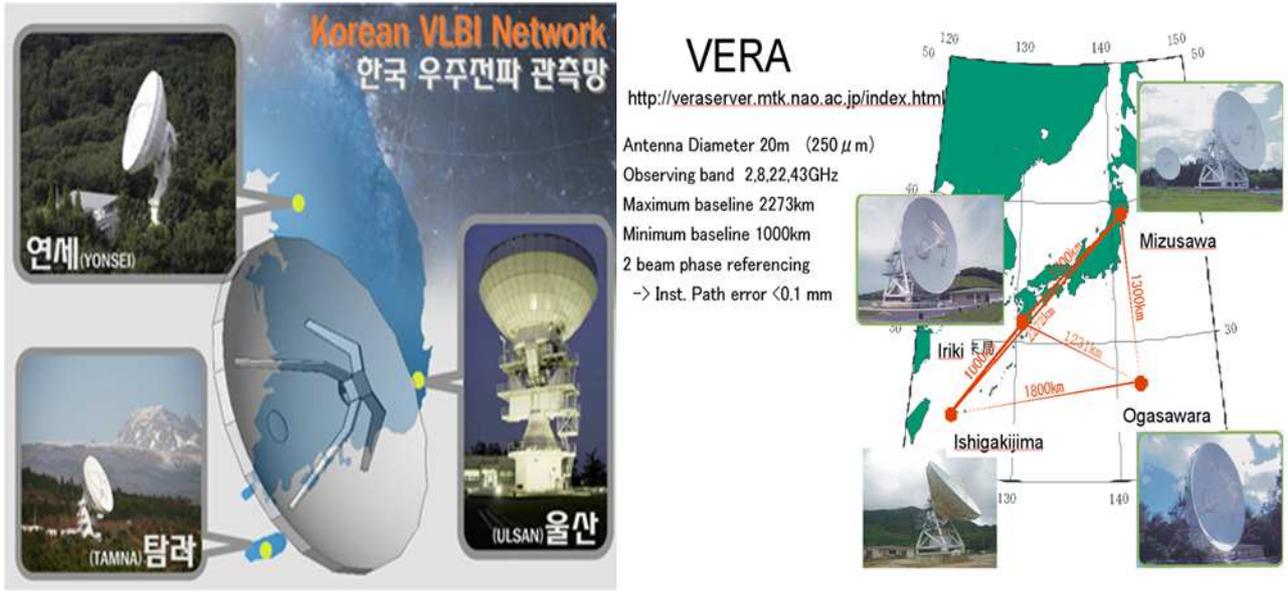}
        \caption{ 
	Korean VLBI Network (KVN) operated by KASI, Korea (left) and
        VLBI Exploration of Radio Astrometry (VERA) operated by NAOJ,
	Japan (right).
        }
        \label{fig2} 
\end{figure*}

A new hardware correlator (the Daejeon correlator, Figure~\ref{fig1}) was developed
in 2006-2009 by the Korea Astronomy and Space Science Institute (KASI) 
and the National Astronomical Observatory of Japan (NAOJ).
The Daejeon correlator will be the main correlator 
for the East Asian VLBI Network (EAVN) consisting of
the Korean VLBI Network (KVN) in Korea~\citep{lee+11,lee+14},
the Japanese VLBI Network (JVN) including  
the VLBI Exploration of Radio Astrometry (VERA) in Japan,
and the Chinese VLBI Network (CVN).
This is one of the main motivations for developing the new hardware
correlator rather than adopting a software correlator
like DiFX~\citep{del+07,del+11}.
The capability of efficient correlation for large number of stations
(e.g., $>$20 for EAVN stations at 22~GHz)
is one of the advantages of the hardware correlator.
In 2010 we started commissioning operations of the correlator,
and began to evaluate its performance using test observations
with the combined network of KVN and VERA (Figure~\ref{fig2}).

The Daejeon hardware correlator
is located in the Korea--Japan Correlation Center (KJCC), Korea.
The Daejeon correlator consists of
several VLBI data playback systems,
a Raw VLBI Data Buffer (RVDB), 
a VLBI Correlation Subsystem (VCS), and
a data archive system.
The Daejeon correlator was aimed to correlate data
obtained from various VLBI networks in East Asia: the KVN,
the KVN and VERA combined network, and the East Asia VLBI Network.
As these use different recording systems,
the Daejeon correlator has several VLBI data playback systems:
Mark 5B, VERA2000, OCTADISK, etc.
They have different interface for data transmission. 
The difference in the formats of the playback system led to 
the introducion of the RVDB system, a big data server with several interfaces.
Data from the RVDB are transferred to the VCS, the main part of the Daejeon
correlator, and correlated with proper control parameters
provided by the correlator control and operation computers.
The VCS is able to process a maximum of 16 stations
with a maximum recording rate of 8192 Mbps, and 8192 output channels for VLBI observation data.
The correlation results from the VCS are then transferred to
the data archive system.
Detailed description of the correlator will be presented
elsewhere (C.~S. Oh et al. 2015, in preparation), and its current status
can be found in the Internet.\footnote{\url{http://kjcc.kasi.re.kr}}

Performance of the new correlator can be evaluated via
investigating the outputs from the whole cycle of the VLBI observations:
data acquisition, correlation, and post-correlation process.
Careful comparison of the output from the new correlator 
with that from an existing, reliable
correlator~\citep[e.g., DiFX as described in][]{del+07,del+11}
should be conducted.
 
We conducted test observations of compact radio sources using KVN and VERA,
correlated the observed data
with the Daejeon correlator and the DiFX
software correlator, and performed careful comparison of the outputs. 
An extensive evaluation using the Daejeon correlator, the DiFX correlator,
and Mitaka FX correlator in NAOJ~\citep{shi+98} 
for the KVN and VERA observations has been conducted.
Full results of the extensive comparison will be reported
elsewhere (C.~S. Oh et al. 2015, in preparation).
In this paper, we report the results of the comparison
between the Daejeon correlator and the DiFX correlator
using the observations only with the KVN.
In Section 2, 
we describe the observations, correlation,
and the post-correlation process.
The results of this comparison are reported
in Section 3,
and we discuss an unusual pattern of the correlation output
in Section 4. We make conclusions
in Section 5.

\section{Observations and Data Reduction\label{sec:obsdata}}

\subsection{Observations Using KVN and VERA\label{sec:obs}}

We observed compact radio sources, 3C~454.3, 3C~345, NRAO~512
and Sagittarius A* (Sgr~A*) at 22~GHz,
on 2011 January 28 UT 19:00 and January 29 UT 04:00,
using KVN and VERA (7 stations):
KVN Yonsei (KY), KVN Ulsan (KU), KVN Tamna (KT),
Mizusawa (MIZ), Ishigakigima (ISH), Ogasawara (OGA),
and Iriki (IRK).
In order to evaluate the various correlation performances of the Daejeon
correlator, the several observing modes and scan lengths were considered.
For this comparison, we used longer scans (10-20 min) for all sources.
The observing frequency is 22.00--22.45~GHz with a selected bandwidth of 256~MHz
and in left circular polarization (LCP).
The received signals within the frequency bandwidth
are 2-bit quantized by AD converters located in the telescope cabin
and transferred to the observing building via optical fibre.
The transferred digital signals are divided into 16 sub-bands (IFs)
by a digital filter bank, and recorded in magnetic disks or tapes by
two data acquisition systems: the Mark 5B system (disk-based) in KVN and
the VERA2000 system (tape-based) in VERA.
The recording rate is 1024~Mbps.

\subsection{Correlation Using Daejeon Correlator and DiFX\label{sec:corr}}

The recording media were sent to the KJCC.
The shipped media were played back using Mark 5B for KVN data
and VERA2000 for VERA data. 
The data from the playback systems were transferred to the RVDB
to be correlated in the VCS.
Finally, the data were correlated using a 1.6384~s correlator
integration time
and 128 spectral channels across each 16~MHz band.
The correlation was performed in 2013 July.
The data recorded at KVN stations were also correlated with
the DiFX software correlator.
Each scan recorded in the Mark 5B disk packs was extracted
by using the utility FUSE,\footnote{\url{http://fuse.sourceforge.net}}
and was transferred to the file server through the 10 Gigabit network.
Scans for strong sources were first used for searching
the clock offset and rate of each station,
and then all scans were correlated.
The correlator integration time was 2.048~s for DiFX
and  the number of spectral channels was 128 across each 16~MHz band.

\subsection{Post-Correlation Process Using AIPS and DIFMAP\label{sec:analysis}}

\begin{figure}[!t]
        \centering 
        \includegraphics[width=80mm]{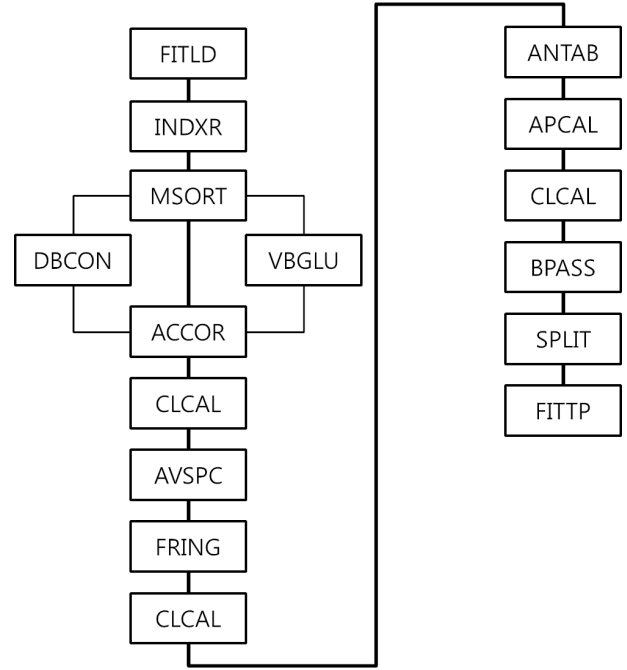}
        \caption{ 
        Schematic diagram of VLBI data reduction with AIPS
        }
        \label{fig3} 
\end{figure}

We performed further processing using AIPS. 
We followed a standard processing procedure as described in Figure~\ref{fig3}.
The correlated output in FITS format
was uploaded into an AIPS catalog by the FITLD task,
and information of the FITS file were indexed with the INDXR task.
In using FITLD, we did not apply digital correction (Digicor = -1),
since it is not necessary for the two correlators. 
We sorted the indexed catalog in an order of time and baseline (TB)
and if necessary, indexed them again.
FITS files from the correlator that were divided into several files in time
were combined using the DBCON task,
while those divided in frequency were combined with the VBGLU task.
At KVN and VERA, we sampled the received signals using the digital samplers.
During the digital sampling, there could be some amplitude errors
of cross-correlation power spectrum.
These errors can be corrected using the ACCOR task based on the amplitude
of the auto-correlation power spectrum for each station. 
These corrections for the amplitudes are stored in a solution (SN) table,
separate from the FITS file itself.
In order to apply the correction in the SN table, we used
the CLCAL task and generated a new calibration (CL) table.
At this stage, one may average the FITS data in frequency
using the AVSPC task. We did not average the FITS data for this work.
The residuals of the fringe delay
and delay rate after the correlation
were searched using the FRING task. This is known as fringe-fitting.
FRING provides us with antenna-based solutions
of the delay and delay rate based on baseline-based solutions.
The solution interval for the fringe-fitting
should be pre-determined according to the coherence time
of the observation at the observing frequency.
For this observation at 22~GHz, we used 1 min as the solution interval
of the fringe-fitting.
Once we obtained the solutions of the delay and delay rate,
we calibrated the phase using CLCAL.
After the phase calibration, we performed the amplitude calibration
in order to correct for the atmospheric opacity change and
for the amplitude errors due to atmospheric fluctuations.
We used system temperatures and antenna gain measured at each observatory
for converting the correlation coefficient to sky brightness and
correcting for the amplitude errors.
We stored the calibration information in the TY and GC tables using
the ANTAB task and produced the solution of the amplitude calibrations
using the APCAL task.
We did not correct for the effect of the bandpass filter
on the spectrum shape in order to investigate the correlation effect
on the spectrum shape.
We divided the FITS data for each source with applying all calibration
information and exported into $uv$-file using FITTP.
For the purpose of the comparison of images,
we used the central 115 channels of each IF, excluding the first 6 and last
7 channels. 

After the phase and amplitude calibration, we made the contour maps
of the target sources using the Caltech DIFMAP software~\citep{she+94}.
We averaged the $uv$-data (or visibility data)
in a time interval of 30~s,
and flagged outliers in amplitude
in order to increase the data processing speed and decrease random error
of individual visibility data.
We should note that the averaging time interval of 30~s is shorter
than the coherence time of this observation.
As a first step of mapping (or imaging) with DIFMAP,
we fitted a point-source model to the visibility
and self-calibrated the phase according to the model,
in order to find the converged model to the visibility.
As a second step, we used the CLEAN deconvolution algorithm
and the amplitude- and phase-self-calibration alternatively.
The CLEAN deconvolution is
a technique to deconvolve the visibility data to find
the true sky brightness distribution of the target source
by establishing a group of delta-function models.
The self-calibration is an algorithm
to reduce the difference between the models and visibility phase/amplitude
using closure phase and closure amplitude.
We should note that the amplitude-self-calibration can not be performed
for the observations with three or less stations.
Therefore, for the analysis of these KVN data, we did not perform
amplitude-self-calibration.
We evaluated the quality of the final map, 
by investigating the residual noise in the image as described
in \cite{lob+06}.
We can quantitatively estimate the noise in the final CLEANed image 
based on the ratio of image noise rms to its mathematical expectation, $\xi_r$.
Suppose that a residual image has an rms $\sigma_{\rm r}$ and 
a maximum absolute flux density $|s_{\rm r}|$. For Gaussian 
noise with a zero mean, the expectation of $s_{\rm r}$ is
\begin{equation}
|s_{\rm {r,exp}}| = \\
\sigma_{\rm r} { \left[ \sqrt{2} \ln{\left( \frac{N_{\rm pix}}
{\sqrt{2\pi}\sigma_{\rm r}} \right) }\right]}^{1/2} ,
\end{equation}
where $N_{\rm pix}$ is the total number of pixels in the image. 
The ratio $\xi_r$ is given by 
\begin{equation}
\xi_{\rm r} = s_{\rm r} / s_{\rm {r,exp}}.
\end{equation}
When the residual noise is similar to Gaussian noise, 
$ \xi_{\rm r} \rightarrow 1 $. If $ \xi_{\rm r} > 1 $,
not all the structure has been adequately cleaned or recovered; 
if $ \xi_{\rm r} < 1 $, the image model obtained has
an excessively large number of degrees of freedom.
In order to compare quantitative information of the final images,
we used circular Gaussian-component models
to fit the self-calibrated data,
yielding the following modelfit parameters:
the total flux density $S_{\rm t}$,
size $d$,
radial distance $r$,
and position angle $\theta$
of each component.

\section{Results\label{sec:results}}

\subsection{Visibility Comparison with DiFX\label{sec:comparision}}

\begin{figure*}[!t]
        \centering 
        \includegraphics[width=170mm]{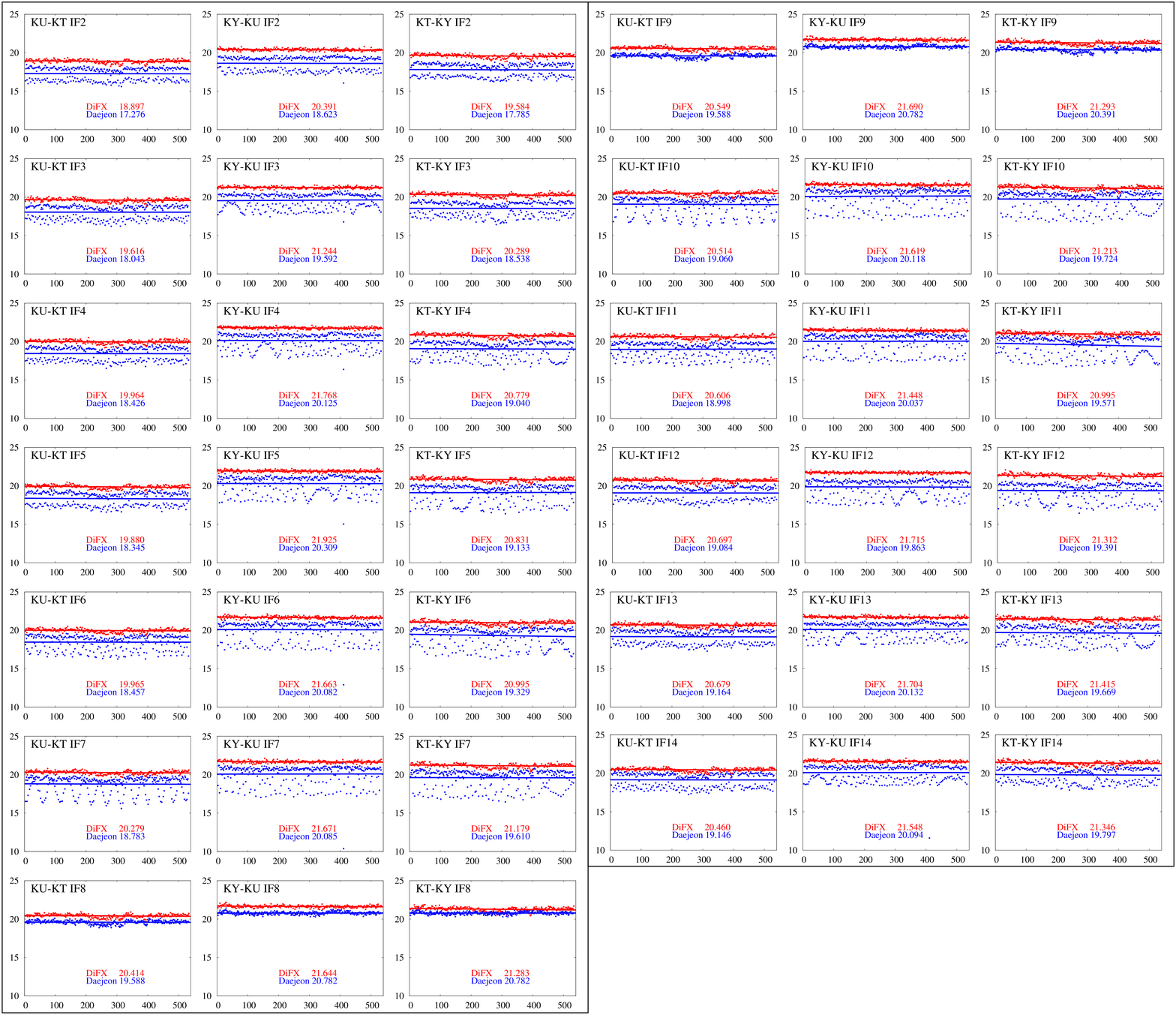}
        \caption{ 
	Comparison of the visibility amplitude as a function of time
	for the source 3C~454.3 as calculated by the DiFX (red dots)
	and the Daejeon correlators (blue dots)
        for IF 2-14 and three KVN baselines: KU-KT, KY-KU, and KT-KY
	during 9 min after 2011 January 29 04:20:00 UT.
	The visibility amplitude is in Jy
	and the time is in seconds from the start of the scan.
	The solid lines represent linear fits to the visibility
	amplitude for the DiFX (red line) and the Deajeon correlator (blue line).
	The numbers on each panel indicate the averaged amplitude
	for each correlator over the time period.
        }
        \label{com-at-3c454.3} 
\end{figure*}

\begin{figure*}[!t]
        \centering 
        \includegraphics[width=170mm]{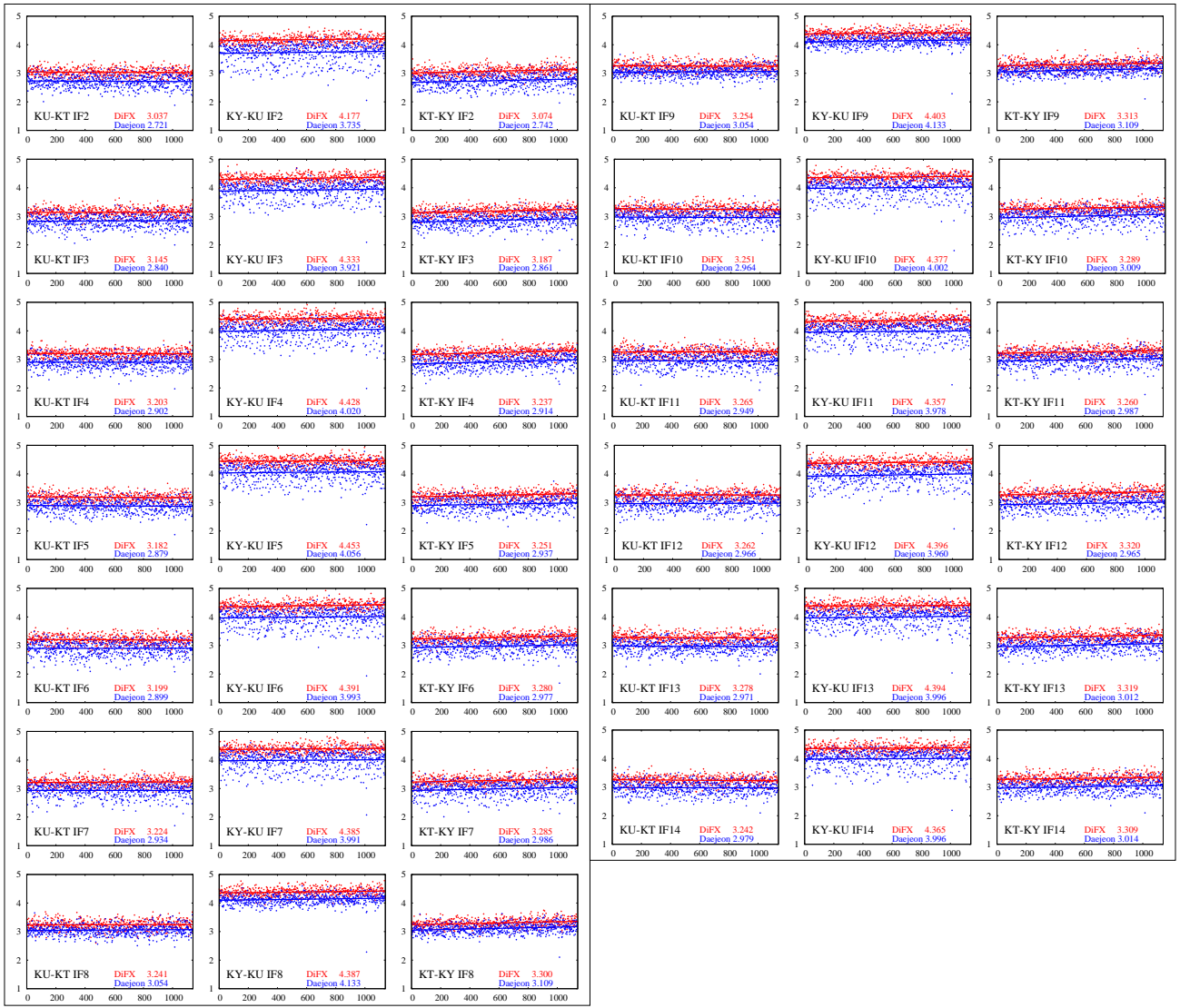}
        \caption{ 
	Same as Figure~\ref{com-at-3c454.3} but for 
	the source 3C~345 and the time range 
	during 19 min after 2011 January 28 19:30:00 UT.
        }
        \label{com-at-3c345} 
\end{figure*}

\begin{figure*}[!t]
        \centering 
        \includegraphics[width=170mm]{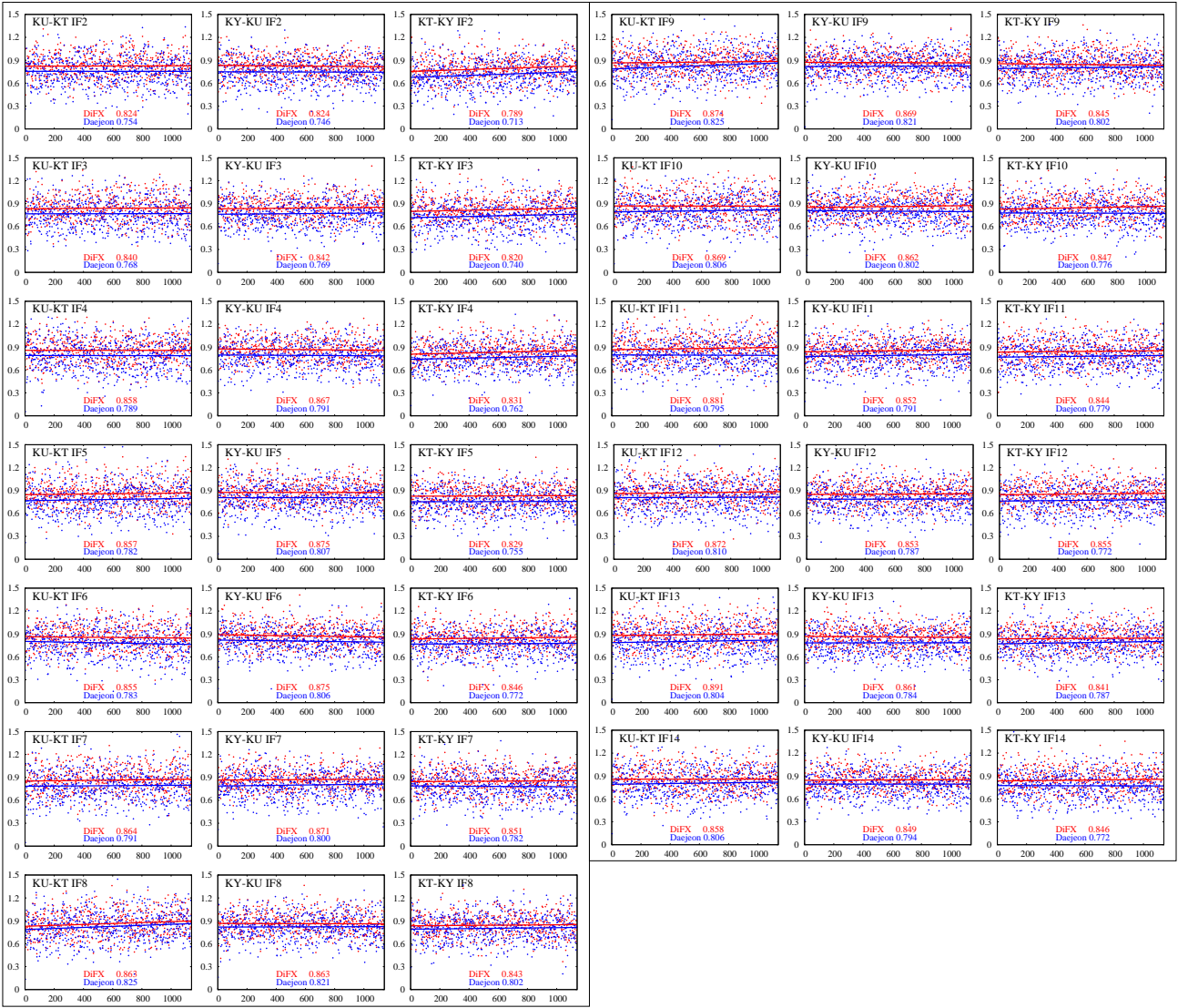}
        \caption{ 
	Same as Figure~\ref{com-at-3c454.3} but for 
	the source NRAO~512 and the time range 
	during 19 min after 2011 January 28 19:10:00 UT.
        }
        \label{com-at-nrao512} 
\end{figure*}

\begin{figure*}[!t]
        \centering 
        \includegraphics[width=170mm]{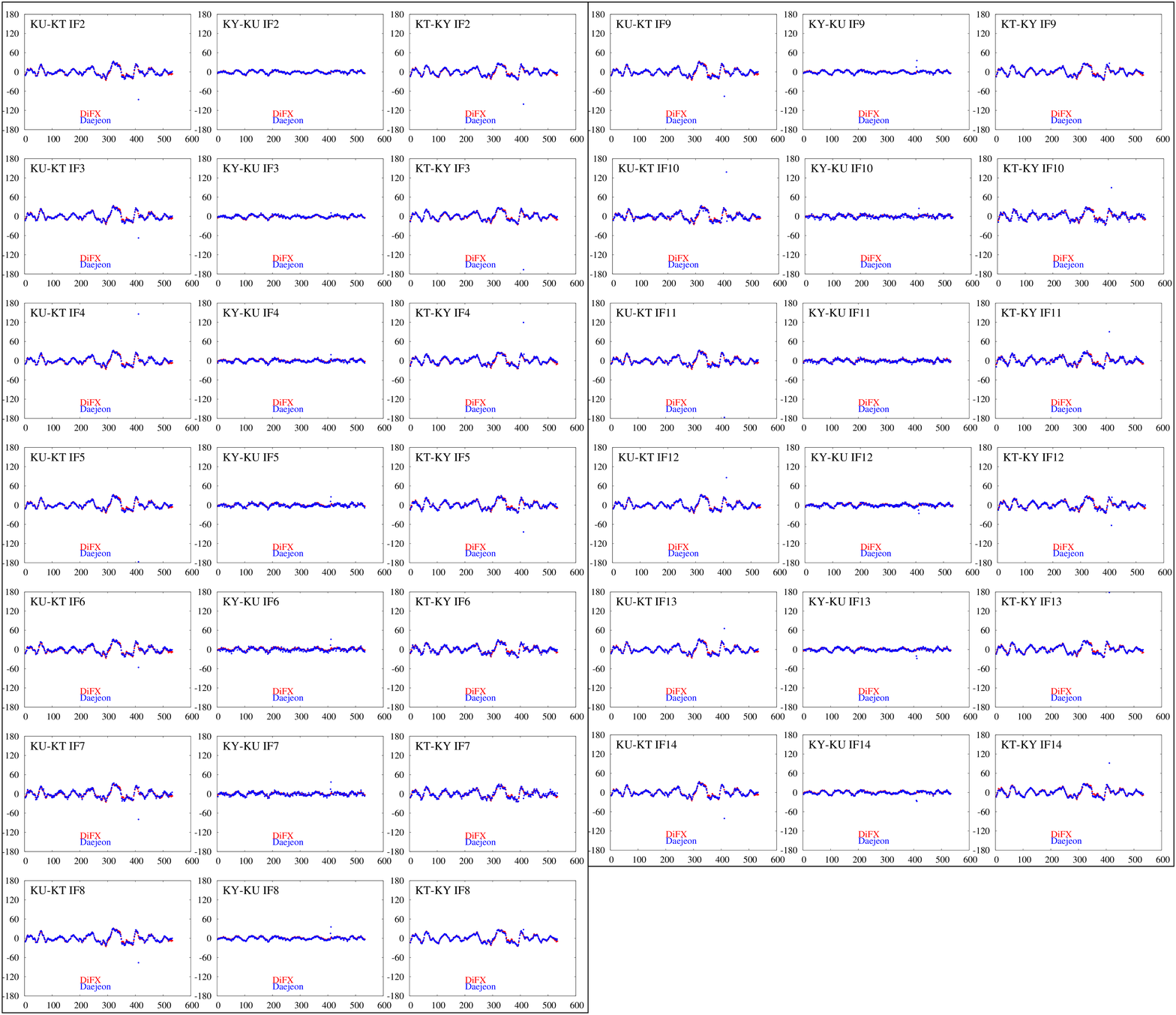}
        \caption{ 
	Comparison of visibility phase as a function of time
	for 3C~454.3 from DiFX (red dots)
	and the Daejeon correlator (blue dots) outputs
        for IF 2-14 and three KVN baselines: KY-KT, KY-KU, and KT-KY
	during 9 min after 2011 January 29 04:20:00 UT.
	The visibility phase is in degrees
	and the time is in seconds from the start of the scan.
        }
        \label{com-pt-3c454.3} 
\end{figure*}

\begin{figure*}[!t]
        \centering 
        \includegraphics[width=170mm]{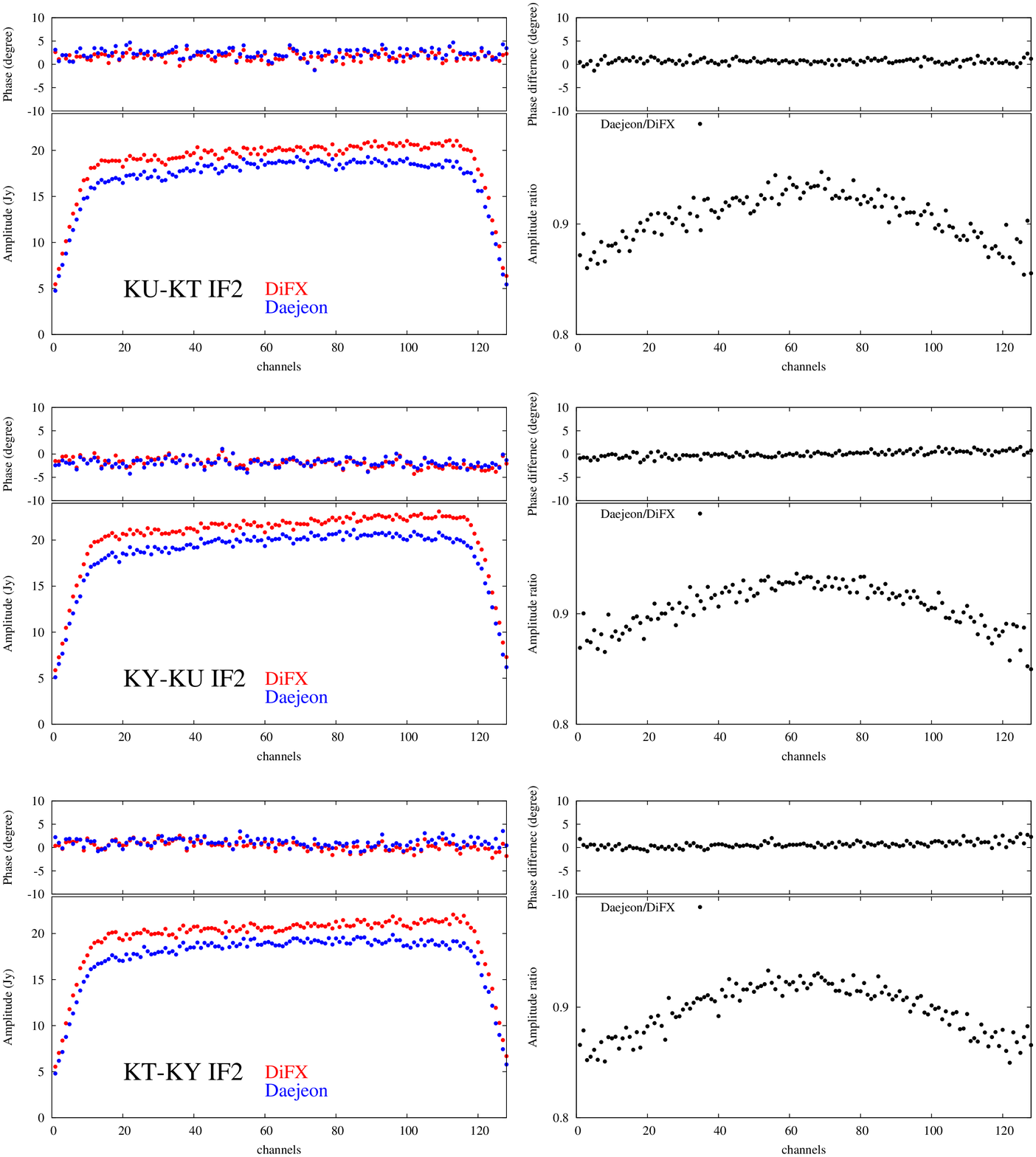}
        \caption{ 
	Left: Comparison of the visibility phase (upper panel of each spectrum)
	and amplitude (lower panel of each spectrum) as a function of frequency
	for 3C~454.3 from the DiFX (red dots)
	and the Daejeon correlator (blue dots) outputs
        for IF 2: 22.034--22.050~GHz band and three KVN baselines:
	KU-KT (top), KY-KU (middle), and KT-KY (bottom)
	during 1 min after 2011 January 29 04:20:00 UT.
	The visibility phase is in degrees spanning a range of $\pm10^{\circ}$
	and the visibility amplitude is in Jy.
	Right: Phase difference (upper) and amplitude ratio (lower)
	for two correlators. The phase is shown with the same scale as the left
	panels while the amplitude ratio spans the range of 0.8--1.0.
        }
        \label{com-possm-3c454.3} 
\end{figure*}

After the calibration and data reduction,
we compared the calibrated output of the Daejeon correlator with
that of DiFX. For a careful comparison, we used the same parameters
and data reduction procedures for the two cases,
except for the correlator integration time
(1.6384~s for Daejeon and 2.048~s for DiFX)
and the delay models
(a Mitaka delay model for Daejeon correlator and
a CALC9 delay model for DiFX).

Figures~\ref{com-at-3c454.3}-\ref{com-at-nrao512} show the visibility
amplitudes on the sources, 3C~454.3, 3C~345, and NRAO~512
for IF 2-14 and three KVN baselines from both correlators
as a function of time. The whole frequency channel (128) data
have been vector-averaged. We fitted the visibility amplitudes
for each baseline
with a first-order polynomial $V(t)=At+B$ using a linear least-squares method.
The fitted models and averaged values are shown in
Figures~\ref{com-at-3c454.3}-\ref{com-at-nrao512}. 
The averaged visibility amplitudes over IF 2-14 within full frequency channels
and central 115 channels for 
individual baselines and sources are summarized in Table~\ref{radplot}.
In Table~\ref{radplot}, we summarized the mean visibility amplitudes
of IF 2-14 for two cases of using full frequency channels in IF
and of using only central 115 channels. 
The visibility amplitudes of the target sources are
in the range of 0.5--22~Jy.
The visibility amplitudes for the Daejeon correlator seem
relatively lower than those for DiFX.
We found that the averaged visibility amplitudes of individual baselines
for the Daejeon correlator are lower by $\leq8\%$ than
those of DiFX for all sources and all baselines.

The amplitude difference between two correlators can be attributed
mainly to an unusual pattern of the visibility amplitude of
the Daejeon correlator,
so called double-layer pattern, as clearly shown
in Figure~\ref{com-at-3c454.3}. 
The double-layer pattern consists of two main patterns
at higher and lower amplitude values. 
One pattern is similar to that of DiFX (normal pattern).
The other has lower values in amplitude than the main pattern
and sometimes is similar to a sinusoidal pattern.
For the baseline KU-KT in IF 2 in Figure~\ref{com-at-3c454.3},
one third of the visibility data
are along the level of about 16~Jy,
and the remaining data are at about 18~Jy.
For the baseline KU-KT in IF 7, one third of the visibility data
show a sinusoidal pattern with its amplitude of about 2~Jy,
that is about 15\% of the flux density.
The double-layer patterns appear different in IF (i.e., in frequency)
and are present in all baselines and IF bands except for IF 8 and 9.
Among the IF 2--14 for the baseline KU-KT,
for IF 2--5 and IF 12--15
the double-layer patterns are shown as two parallel patterns,
and for IF 6--7 and IF 10--11
the double-layer patterns appear like the sinusoidal pattern.
The visibility amplitudes for IF 8 and 9 do not show
such a double-layer pattern.
It seems that the period of the sinusoidal pattern varies in time
and frequency. The periods for IF 6 and 11 seems
twice smaller than those for IF 7 and 10, respectively.
We found also that there is a rough symmetric trend
of the double-layer patterns between IF 2--8 and IF 9-15.
This implies that the double-layer patterns
may happen in a correlation stage related with frequency.
The double-layer patterns are clearer for the observations
of stronger sources like 3C~454.3 with a correlated flux density
of $\sim$20~Jy (Figure~\ref{com-at-3c454.3}), whereas they are less clear
for relatively weaker sources like 3C~345 and NRAO~512
with flux densities of 1--4~Jy (Figures~\ref{com-at-3c345} and \ref{com-at-nrao512}).
This is because the amplitude of the double-layer pattern
is about 15\% and it is similar to 
the amplitude calibration uncertainty of KVN observations
at 22~GHz~\citep[see][]{pet+12,lee+14}.
However, the 8\%-difference of visibility amplitudes between 
DiFX and the Daejeon correlator appears in all sources.

Figure~\ref{com-pt-3c454.3} shows the visibility
phases of the source 3C~454.3
for IF 2-14 and three KVN baselines from both correlators
as a function of time. The whole frequency channel (128) data
have been vector-averaged.
We found that the visibility phases show small differences due to
the different delay models used for the two correlators.

Figure~\ref{com-possm-3c454.3} shows a comparison
of the cross-correlated spectrum,
i.e., the visibility amplitude and phases as a function of frequency
in one 16~MHz band, IF 2. The spectra have been
vector-averaged over one minute time range.
The phases of the cross-correlated spectrum are
consistent with each other within $\leq$2 degrees,
whereas the amplitudes are different by about 10\% on average.
The amplitude difference varies in frequency.
At the band edges, the difference gets larger, whereas at the band center,
it becomes a bit smaller. This indicates that
the spectrum shape of Daejeon correlator output is a bit different
from that of DiFX.  

\begin{table*}[t!]
\caption{Comparison of visibility amplitudes\label{radplot}}
\centering
\resizebox{175mm}{!}{
\begin{tabular}{llcccccccc}
\toprule
       &            & \multicolumn{2}{c}{$S_{\rm KU-KT}$} && \multicolumn{2}{c}{$S_{\rm KY-KU}$} && \multicolumn{2}{c}{$S_{\rm KT-KY}$} \\
\cline{3-4} \cline{6-7} \cline{9-10} \addlinespace
Source & Correlator & FULL BW & 115CH && FULL BW & 115CH && FULL BW& 115CH \\
(1) & (2) & (3) & (4) && (5) & (6) && (7) & (8) \\
\midrule
3C~454.3 & DiFX    & $20.2\pm0.547$ & $21.2\pm0.576$ & & $21.5\pm0.392$ & $22.6\pm0.413$ & & $21.0\pm0.554$ & $22.0\pm0.586$ \\ 
         & Daejeon & $18.8\pm1.04$  & $19.8\pm1.09$  & & $20.0\pm1.07$  & $21.1\pm1.13$  & & $19.4\pm1.17$  & $20.4\pm1.23$  \\ 
         & $S_{\rm Daejeon}/S_{\rm DiFX}$ & 0.93(1.9)& 0.93(1.9) & & 0.93(2.7) & 0.93(2.7) &&   0.92(2.1) &	0.93(2.1)\\ \addlinespace 
3C~345   & DiFX    & $3.21\pm0.169$ & $3.37\pm0.181$ &&  $4.37\pm0.158$ & $4.59\pm0.170$ & & $3.26\pm0.167$ & $3.43\pm0.179$ \\ 
         & Daejeon & $2.93\pm0.236$ & $3.09\pm0.252$ &&  $3.99\pm0.291$ & $4.21\pm0.308$ & & $2.97\pm0.239$ & $3.13\pm0.255$   \\ 
         & $S_{\rm Daejeon}/S_{\rm DiFX}$ & 0.91(1.3)&	0.92(1.4)&&  0.91(1.84)&	0.92(1.8) & &0.91(1.4)&	0.91(1.4) \\  \addlinespace

NRAO~512 & DiFX    & $0.861\pm0.163$ & $0.908\pm0.176$ & & $0.858\pm0.140$ & $0.903\pm0.153$ & & $0.837\pm0.149$ & $0.881\pm0.161$   \\ 
         & Daejeon & $0.793\pm0.188$ & $0.839\pm0.203$ & & $0.792\pm0.169$ & $0.837\pm0.183$ & & $0.768\pm0.172$ & $0.812\pm0.186$  \\ 
         & $S_{\rm Daejeon}/S_{\rm DiFX}$ & 0.92(1.2)&	0.92(1.2)&  &0.92(1.2)	&0.93(1.2)&  &0.92(1.2)	&0.92(1.2)\\ 
\bottomrule
\end{tabular}
}
\vskip6pt
\tabnote{ 
Column designation: 
1~-~source name;  
2~-~correlator;  
3-9~-~mean of visibility amplitude for KU-KT, KY-KU,
and KT-KY in Jy,
or the ratio of the mean values for Daejeon to those for DiFX
with the ratio of their standard deviations in parentheses:
Columns 3,5,7 for values using data in full bandwidth
and Columns 4,6,8 for using data excluding first 6 channels and last 7 channels in each band (hence averaged over 115 channels).    
}
\end{table*}

\subsection{Image Comparison with DiFX\label{sec:comparision-2}}

\begin{figure*}[p]
        \centering 
        \includegraphics[width=175mm]{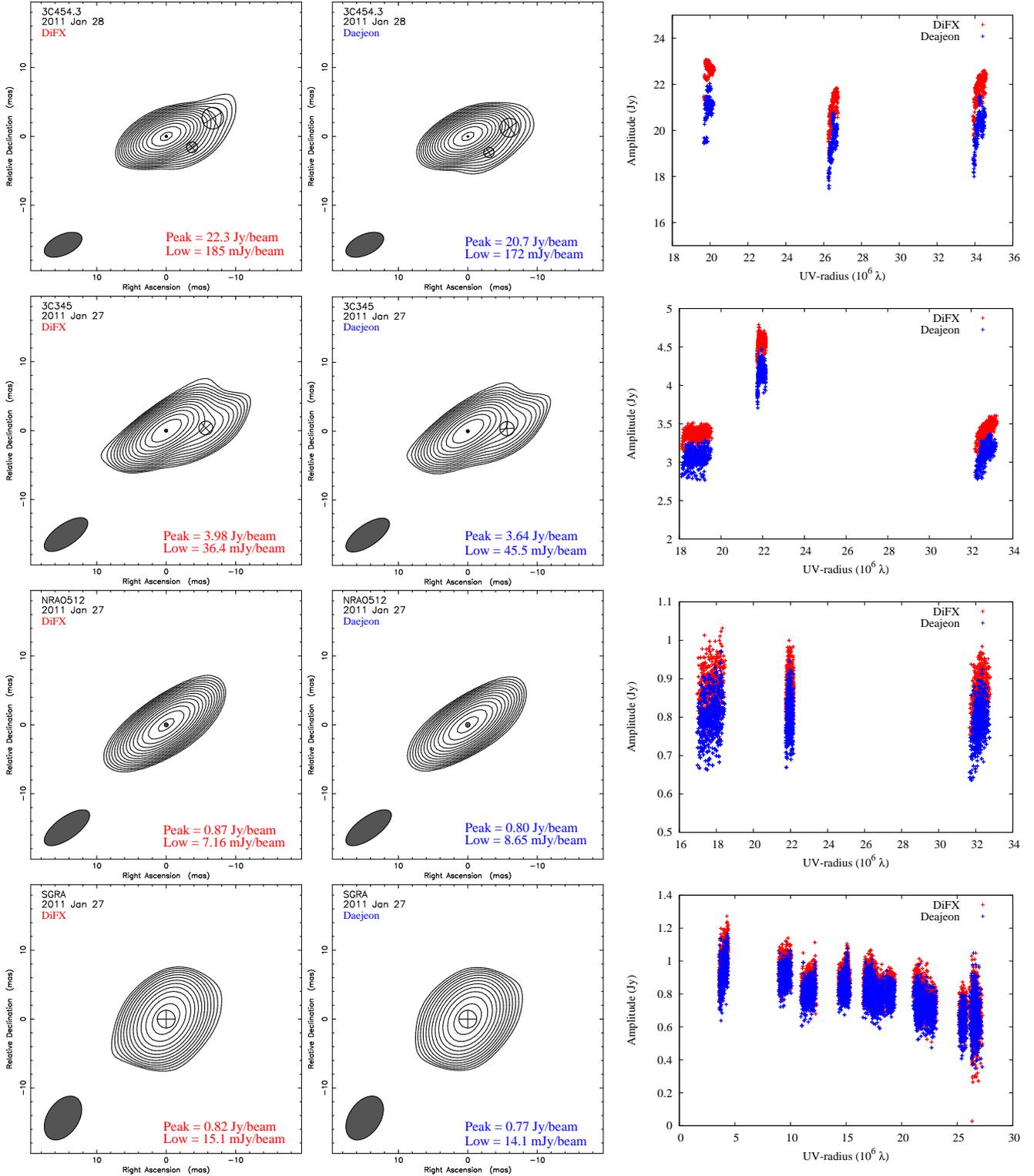}
        \caption{ 
	Left: CLEANed images using the DiFX correlator outputs.
	Circular Gaussian models are on top of the contour maps.
	The axes of each map are the relative R.A. and decl. offsets
	from the tracking center in milliarcseconds.
	The lowest contour level is shown in the lower right corner of
	each map. The contours have a logarithmic spacing and
	are drawn at 1, 1.4, ..., $1.4^n$ of the lowest contour level.
	Middle: CLEANed images using the Daejeon correlator outputs
	in the same format as the left panel.
	Right: distributions of the visibility amplitude for DiFX (red)
	and the Daejeon correlator (blue) against the $uv$ radius.
	The $x$ axis shows the $uv$ distance in $10^6\lambda$,
	and the $y$ axis represents the visibility amplitude
	(correlated flux density) in Jy, averaged over 30 s.
	Image parameters of each image are summarized in Table~\ref{imagepar}.
        }\label{fig-map} 
\end{figure*}

In addition to the comparison of the visibility amplitudes and phases
in time and frequency,
we compared the imaging results from the two correlator outputs
for all target sources, 3C~454.3, 3C~345, NRAO~512, and SgrA$^*$.
As mentioned previously, we used the same procedure of imaging
for the comparison: (a) fitting a point source to the visibility,
(b) CLEAN and phase-self-calibration, and (c) modelfit with
circular Gaussian components.
In Figure~\ref{fig-map}, we compare CLEANed images 
for the DiFX and Daejeon correlators.
We show the contour maps of each source for DiFX in the left panel,
and those for Daejeon correlator in the right panel.
There are circular Gaussian models on top of the contour maps.
The $x$ and $y$ axes are in units of mas.
The sources are 3C~454.3, 3C~345, NRAO~512, and SgrA$^*$ from top to bottom.
For each contour map, the source name and the observation date
are given in the upper left corner of the map.
In the lower right corner of the map, we show
the peak flux density and the lowest contour level in units of mJy.
The shaded ellipse in the lower left corner of the map
represents the FWHM of the restoring beam of the image.
In all of the images, the contours are drawn at 1, 1.4, ..., $1.4^n$
(logarithmic spacing) of the lowest flux density level.
In the right panel, we showed the visibility amplitudes 
as a function of the $uv$ radius.
The $x$ axis of the plot represents the $uv$ radius in units of $10^6\lambda$,
where $\lambda$ is the observing wavelength,
which is the length of the projected baseline used to obtain
the visibility data.
The $y$ axis of the plot shows the amplitude of each visibility point
(i.e., correlated flux density) in units of Jy.
The quality of the final images was investigated
by estimating the quality of the residual noise, $\xi$,
as described in Section~\ref{sec:analysis}.
More detailed parameters of the images presented in Figure~\ref{fig-map}
are summarized in Table~\ref{imagepar}.
For each image, Table~\ref{imagepar} lists
the source name, the parameters of the restoring beam
(the size of the major axis, $B_{\rm maj}$,
the minor axis, $B_{\rm min}$, and
the position angle of the beam, $B_{\rm PA}$),
the total flux density, $S_{\rm t}$,
the peak flux density, $S_{\rm p}$,
the off-source rms, $\sigma$,
the dynamic range of the image, $D$,
and the quality of the residual noise, $\xi$, for each image.
The image results show that the total flux and peak flux densities
for all sources are consistent within 8\% for the two correlator outputs.
The difference is similar to that of the visibility amplitude averaged
for each baseline as summarized in Table~\ref{radplot}.
In order to investigate the flux differene in more detail,
we made images using data for IF 8 and 9, for which the double-layer pattern
does not appear in Figure~\ref{com-at-3c454.3}.
We compared the total and peak flux density for the images free of the double-layer-pattern,
and found that their difference is on average about 5\% for all sources, 
as summarized in Columns (11) and (12) of Table~\ref{imagepar}.
This may indicate that a part of the total difference of 8\% is caused by
other differences between the two correlators that are not related with the double-layer pattern.
We also compared the flux density, size, and position of
the core and jet components of all sources
using circular Gaussian model-fitting.
Table~\ref{modelpar} lists the parameters of each model-fit
component: the total flux, $S_{\rm t}$,
size, $d$,
angular distance from the central component, $r$,
and position angle, $\theta$
(the location of the jet component with respect to the core component).
The modelfit parameters show that the total flux densities
of core and jet components of almost all sources are
again consistent within 8\% for the two correlator outputs,
whereas the sizes of the Gaussian components are different by $\le40$\%.
For 3C~454.3 and 3C~345, we were able to modelfit the jet components for the two correlator
outputs with very consistent positions to each other.

\begin{table*}[t!]
\caption{Comparison of imaging parameters\label{imagepar}}
\centering
\begin{tabular}{llcccccccccc}
\toprule
Source & Correlator & $B_{\rm maj}$ & $B_{\rm min}$ & $B_{\rm PA}$ & $S_{\rm t}$ & $S_{\rm p}$ & $\sigma$ & D & $\xi_{r}$ & $S_{\rm t,IF8-9}$ & $S_{\rm p,IF8-9}$ \\
(1) & (2) & (3) & (4) & (5) & (6) & (7) & (8) & (9) & (10) & (11) & (12)\\
\midrule
3C~454.3 & DiFX    & 5.829 & 2.937 & -65.1 & 24.10        & 22.27       & 61.53  & 362 & 0.64 &24.17  &22.28\\ 
         & Daejeon & 5.835 & 2.939 & -65.1 & 22.26        & 20.71       & 57.12  & 363 & 0.76 &23.23  &21.41\\
         & $Daejeon/DiFX$ &-&-     & -     & 0.92         & 0.93        & 0.93   & 1.0 & -    &0.96   &0.96 \\ \addlinespace
3C~345   & DiFX    & 7.323 & 3.111 & -55.1 & 4.703        & 3.978       & 12.14  & 328 & 0.57 &4.656  &4.018 \\
         & Daejeon & 7.333 & 3.114 & -55.1 & 4.321        & 3.640       & 15.17  & 240 & 0.56 &4.390  &3.790 \\
         & $Daejeon/DiFX$ &-&-     & -     & 0.92         & 0.92        & 1.2    & 0.73& -    &0.94   &0.94\\ \addlinespace
NRAO~512 & DiFX    & 7.741 & 3.151 & -54.1 & 0.887        & 0.871       & 2.387  & 365 & 0.53 &0.897  &0.876\\ 
         & Daejeon & 7.746 & 3.153 & -54.1 & 0.818        & 0.797       & 2.884  & 276 & 0.53 &0.855  &0.829\\
         & $Daejeon/DiFX$ &-&-     & -     & 0.92         & 0.92        & 1.2    & 0.76& -    &0.95   &0.95 \\ \addlinespace
Sgr~${\rm A}^{*}$ & 
         DiFX      & 7.033 & 4.493 & -33.5 & 1.011        & 0.822       & 5.023  & 164 & 0.88 &1.019  &0.828\\ 
         & Daejeon & 7.042 & 4.494 & -33.6 & 0.946        & 0.772       & 4.715  & 164 & 0.80 &0.988  &0.801\\ 
         & $Daejeon/DiFX$ &-&-     & -     & 0.94         & 0.94        & 0.94   & 1.0 & -    &0.97   &0.97 \\ 
\bottomrule
\end{tabular}
\tabnote{ 
Column designation: 
1~-~source name;  
2~-~correlator;  
3~-~major axis[mas];
4~-~minor axis[mas];
5~-~position angle of the major axis [$^{\circ}$] of the restoring beam;
6~-~total flux density [Jy];
7~-~peak flux density [Jy~beam$^{-1}$],
and the ratio of the total and peak flux densities of Daejeon correlator
to those of DiFX; 
8~-~off-source RMS in the image [Jy~beam$^{-1}$];
9~-~Dynamic range of the image ($D=S_{\rm p}/\sigma$);
10~-~quality of the residual noise in the image;
11-12~-~total flux density [Jy] and peak flux density [Jy~beam$^{-1}$]
for images using data for IF 8 and 9.
}
\end{table*}

\begin{table*}[t!]
\caption{Comparison of model fit parameters\label{modelpar}}
\centering
\begin{tabular}{lcccccccccc}
\toprule
 & \multicolumn{2}{c}{$S_{\rm t}$} && \multicolumn{2}{c}{$d$} && \multicolumn{2}{c}{$r$} & \multicolumn{2}{c}{$\theta$} \\
\cline{2-3} \cline{5-6} \cline{8-9} \cline{10-11} \addlinespace

Source & DiFX & Daejeon & $\frac{S_{\rm Daejeon}}{S_{\rm DiFX}}$ & DiFX & Daejeon &$\frac{d_{Daejeon}}{d_{\rm DiFX}}$& DiFX & Daejeon & DiFX & Daejeon\\
(1) & (2) & (3) & (4) & (5) & (6) & (7) & (8) & (9) & (10) & (11)\\
\midrule
3C~454.3 &  22.4   &  20.8   &0.93& 0.315 & 0.199&0.63 & -    & - & -  & - \\ 
         &  0.872  &  0.835  &0.96& 1.59  & 1.48 &0.93 & 4.05 & 3.83 & -113 & -127\\ 
         &  0.780  &  0.624  &0.80& 3.11  & 2.73 &0.88 & 7.21 & 6.14 & -68.5& -78.2\\ \addlinespace
3C~345   &  4.00   &  3.67   &0.92 & 0.421  & 0.406  &0.96 & -   & -    & -     & - \\
         &  0.702  &  0.648  &0.92 & 2.00   & 2.07   &1.0  & 5.72& 5.66 & -86.2 & -86.4\\\addlinespace
NRAO~512 &  0.887  &  0.818  &0.92 & 0.537 & 0.628  &1.2 & - & - & - & - \\ \addlinespace
Sgr~${\rm A}^{*}$ & 
         1.01   &  0.946  &0.94 & 2.58 & 2.56   &0.99 & - & -& - & - \\
\bottomrule
\end{tabular}
\tabnote{ 
Column designation: 
1~-~source name;  
2-3~-~model flux density of the component[Jy];
4~-~ratio of the model flux density of Daejeon correlator to that of the DiFX;
5-6~-~size of the component[mas];
7~-~ratio of the model size of the Daejeon correlator to that of DiFX;
8-9~-~angular distance from the central component[mas];
10-11~-~position angle[$^{\circ}$];
}
\end{table*}

\section{Discussion\label{sec:discusion}}

\subsection{Double-Layer Patterns\label{sec:discusion-dp}}

The most prominent, apparent difference between correlation outputs
from the Daejeon and DiFX correlators is the double-layer patterns as shown in
Figure~\ref{com-at-3c454.3}.
The double-layer patterns vary in IF (i.e., in frequency)
and are present in almost all baselines including KVN (and also VERA stations).
When we investigated the visibility amplitude of 3C~454.3 for all baselines of
KVN and VERA, we found that
the amplitudes for all baselines show the double-layer patterns,
except for baselines with larger scatter (e.g., Ishigaki-KVN).
For VERA baselines, the double-layer patterns consist of two parallel
patterns, and for KVN baselines we see clearly sinusoidal patterns, as discussed above.
This is consistent with the results for the KVN baselines of fainter sources.

It is important to investigate possible reasons causing the double-layer pattern
and to improve the performance of the Daejeon correlator.
A detailed investigation of the reasons behind the double-layer pattern
is underway by the KJCC engineering team and will be reported elsewhere.
Here we offer a brief discussion of possible ways this pattern can be produced.
Possible reasons include the following:
\begin{itemize} 
\item{\bf Fringe rotation error}.
The plot of the visibility as a function of time
shows very periodic patterns, which may indicate 
some fundamental problem in the correlator: for example,
the delay shifter or the phase rotator is not updated
as it should. However, as the pattern appears
only in the amplitude, it may not be the case. 
Moreover, same patterns are seen on two separate
channels: IF3=IF14, IF4=IF13, IF5=IF12, IF6=IF11,IF7=IF10.
IF8 and IF9 are both OK, IF2 has no match, and IF1 is not plotted.
Since these matching channels have different frequencies,
the fringe rotator is not the problem.
\item{\bf Timing problem between normalization and accumulation}.
By looking at the plot more closely, we found that
some integration intervals may have lost part of the data.
The integration time of 1.6384~s for the Daejeon correlator
is so unusual that may modulate with other time in the correlator system.
Integer number of seconds of integration time, or shorter intervals,
would give better results.
It is possible that the IF pattern mentioned above can be compared with the order in
which the control computer gets the data from the hardware.
We found that this problem does not affect the correlation ouput of DiFX
since there is no data loss on the normalization and accumulation
in DiFX with the time interval of 2.048~s.
\item{\bf Problem with the geometry applied in the correlator}.
The period of the sinusoidal pattern of the double-layer patterns
seems to vary in time and in frequency. This could indicate
that there is a problem with the geometry
applied in the correlator.
The correlation output of VLBI observations with a finite 
bandwidth $\Delta\nu$ is 
$r\propto \frac{sin\pi\Delta\nu\tau_g}{\pi\Delta\nu\tau_g}
cos(2\pi\nu_0\tau_g-\phi)$,
where $\tau_g$ is the geometric delay and
$\nu_0$ is the observing frequency~\citep[see e.g.,][]{tho99}.
The amplitude of the correlation output is modulated by a sinc-function
envelope as $\frac{sin\pi\Delta\nu\tau_g}{\pi\Delta\nu\tau_g}$. Usually the modulation
is well modeled by the geometric model used during the correlation.
However, if there is an error in geometry, i.e., $\Delta\tau = \tau_g-\tau_m$,
where $\tau_m$ is the geometric time delay of the geometric model, then 
we may expect a periodic pattern in the amplitude of the correlation output
equal to $\frac{sin\pi\Delta\nu\Delta\tau_g}{\pi\Delta\nu\Delta\tau_g}$.
If this is the case, the sinusoidal pattern may correlate
with the baseline length. However it is hard to tell
wether we see the correlation in the KVN observations,
since the baseline length of KVN is in the range of 305-477~km.
Moreover, the geometric problem should affect the phase of the correlation
output as $cos(2\pi\nu_0\tau_g-\phi)$. Since we see
no prominent difference in phases between Daejeon and DiFX correlators,
we can exclude this possibility.
\end{itemize}

Although the amplitude of the double-layer patterns is as large as 10-15\%
the final effect to the flux density in the CLEANed image is less than 8\% (maybe 3\%, see below) because the visibility data are averaged in time and frequency
for the final imaging.
However this effect should be investigated more carefully with further test
observations in various observing modes (e.g., full track imaging mode).
Since the comparison in this paper used the observations with
short integration times of 10-20 min (75min for SgrA$^*$),
we have an uncertainty in imaging results.
The image quality values of $\xi_{\rm r} $ of the final images
are in the range of 0.53-0.88.

\subsection{Fringe Tracking\label{sec:discusion-dp-2}}

We found that part (5\%) of the total difference of 8\% in flux density between the two correlators
are caused by other reasons which are not related with the double-layer pattern. 
\cite{igu+00} investigated the performance of
the hardware correlator developed for
the VLBI Space Observatory Programme (VSOP)
and found that the loss of visibility amplitude due to the hardware phase tracking
is 4\%. The loss almost corresponds to the 5\% difference in flux density
of images using the double-layer-pattern free data (i.e., data for IF 8 and 9). 
For the Daejeon correlator, the fringe tracking is done in VCS
in the same way as described in \cite{igu+00}.
So a major part of the 5\% degradation in hardware correlation
may come from the difference in the way of fringe phase tracking.
The additional 1\% of the difference can be attributed to other reasons (e.g., bit-jump correction).

\section{Conclusions\label{sec:conclusion}}

We evaluated the performance of a new hardware correlator
in Korea, the Daejeon correlator,
by conducting KVN VLBI observations at 22~GHz.
We correlated the acquired data with DiFX and the Daejeon correlator
for comparing the visibility data and imaging results.
The flux densities and brightness distributions of the
target sources for the two correlators are consistent with
each other within $<8\%$.
The difference is comparable with the amplitude calibration uncertainties of
KVN observations at 22~GHz.
We also found that the 8\% difference in flux density is caused mainly by
(a) the difference in the way of fringe phase tracking
between the DiFX software correlator and the Daejeon hardware correlator,
and (b) an unusual pattern (a double-layer pattern) in the amplitude 
correlation output from the Daejeon correlator. 
The visibility amplitude loss by the double-layer pattern is as small as 3\%. 
We conclude that the new hardware correlator produces 
reasonable correlation outputs
for continuum
observations
which are consistent with the outputs from the software correlator, DiFX.


\acknowledgments

We thank Walter Alef, David Graham, and Alan Whitney for their
fruitful discussion on the performance of the Daejeon correlator.
We would like to thank the anonymous referee for important comments
and suggestions which have enormously improved the manuscript.
We are grateful to all staff members in KVN
who helped to operate the array and to correlate the data.
The KVN is a facility operated by
the Korea Astronomy and Space Science Institute.
The KVN operations are supported
by KREONET (Korea Research Environment Open NETwork)
which is managed and operated
by KISTI (Korea Institute of Science and Technology Information).

\end{document}